\documentclass[preprint,showpacs,preprintnumbers,amsmath,amssymb]{revtex4}
\usepackage{graphicx}% Include figure files
\usepackage{dcolumn}% Align table columns on decimal point
\usepackage{bm}% bold math

\catcode`\@=11
\def\lsim{\mathrel{\mathpalette\@versim<}}
\def\gsim{\mathrel{\mathpalette\@versim>}}
\def\@versim#1#2{\vcenter{\offinterlineskip
\ialign{$\m@th#1\hfil##\hfil$\crcr#2\crcr\sim\crcr } }}
\catcode`\@=12

\newcommand{\nn}{\nonumber}

\newcommand{\be}{\begin{equation}}
\newcommand{\ee}{\end{equation}}
\newcommand{\bea}{\begin{eqnarray}}
\newcommand{\eea}{\end{eqnarray}}

\newcommand{\psla}{p\hspace{-5pt}/}

\begin{document}
\input epsf.tex

%\tightenlines
%\draft

\preprint{KANAZAWA-07-04}
\title{
Conformal dynamics in gauge theories via
non-perturbative renormalization group
}
\author{Haruhiko Terao}
\email{terao@hep.s.kanazawa-u.ac.jp}
\author{Akito Tsuchiya}
\email{tsucchi-@hep.s.kanazawa-u.ac.jp}
\affiliation{
Institute for Theoretical Physics, Kanazawa
University, Kanazawa 920-1192, Japan
}%
\date{\today}
\begin{abstract}
The dynamics at the IR fixed point realized in the
$SU(N_c)$ gauge theories with massless Dirac fermions
is studied by means of the non-perturbative 
renormalization group.
The analysis includes the IR fixed points with non-trivial
Yukawa couplings.
The renormalization properties of the 
scalar field are also discussed and it is
shown that  hierarchical mass scale may be
allowed without intense fine-tuning 
due to a large anomalous dimension.

\end{abstract}


%\narrowtext

\maketitle

\section{Introduction}

The hierarchy between the Electro-Weak (EW) scale and
the fundamental scales, Planck scale or the grand
unification scale,  has been
a long-standing problem in particle physics.
Contrary to this, smallness of the QCD scale may 
be understood as a consequence of the classical
scale invariance of QCD,
although the scale invariance is broken quantum
mechanically.
The QCD scale is determined by the so-called
dimensional transmutation.
However the fundamental scalar theory necessarily
allows a relevant operator, the scalar mass term,
which causes the gauge hierarchy problem.
Indeed the mass term can be forbidden by some
symmetries such as supersymmetry.
Meanwhile, the relevant operators may be also
suppressed, if the dynamics respects the
conformal invariance.
Therefore, it would be worth while to think about
the conformal field theories
as another possibility to 
shed light on the hierarchy problem.

Recently, conformal field theories (CFTs) in four
dimensions have been also attracting much attention
in the context of the so-called AdS/CFT 
correspondence \cite{AdS/CFT,AdS/CFT2}.
Especially, some phenomenological models
given in the five dimensional warped space
have been studied intensively as new scenarios 
for the  EW symmetry breaking.
The typical ones are the Higgsless models,
the minimal composite Higgs models and so on
\cite{AdS/CFTmodels}.

It would be very interesting to find the
four-dimensional descriptions of the models
given in the five dimensional AdS space.
However the AdS/CFT correspondence is rather
speculative.
Usually explicit CFTs equivalent to the five 
dimensional models are unknown,
although some models are thought to
correspond with the walking Technicolor (TC) 
models \cite{walkingTC}.

On the other hand, 
the explicit examples of the non-trivial CFTs 
are very limited in four dimensions, specially
in the non-supersymmetric cases.
Indeed it has been well-known for some time that 
the non-abelian gauge theory becomes a CFT for 
the appropriate number of flavors.
The gauge coupling approaches the so-called 
Banks-Zaks (BZ) fixed point \cite{BZ} towards
infrared (IR) direction.
However, this class of gauge theories would be 
the unique known example of the CFTs. 

In the Ref.~\cite{BZ}, the existence of the fixed point
was discussed by the zero points of the gauge 
beta function evaluated perturbatively.
Therefore the analysis of the fixed point
is reliable only in the weak coupling region.
However, the gauge coupling of the BZ fixed point 
increases rapidly as the number of flavors decreases. 
Therefore the flavor numbers are found to be
rather limited so as to realize the BZ fixed
points in the weak coupling region.
Then some non-perturbative analysis is required to
clarify the phase diagram for various numbers of 
flavors.

In this paper we examine the BZ fixed point and
the dynamical properties of the CFT even in the
strong coupling region by means of the non-perturbative
renormalization group (NPRG). 
So far the phase diagrams of the $SU(N_c)$ gauge 
theories with various numbers of vector-like fermions 
have been investigated by solving the 
Dyson-Schwinger (DS) equations \cite{DS,ATW,MY}.
In these analyses, the two-loop gauge beta
function was applied in the approximation scheme. 
Then it was shown the flavor number $N_f$
should be larger than $4 N_c$ in the large $N_c$ leading
in order to have the unbroken 
phase of chiral symmetry.
The BZ fixed point should exist in this unbroken
phase, since the theory on the fixed point is
scale invariant.
Thus the conformal window of QCD like theories
is given by $16 \geq N_f \geq 12$,
while numerical simulation of $SU(3)$
lattice gauge theories indicates the window of
$16 \geq N_f \geq 7$ \cite{lattice}.

In the DS approach, the ladder approximation is
frequently applied due to it's simplicity.
It has been known that the exactly same results
of the chiral phase boundary and the order
parameter can be obtained by solving the
NPRG equations \cite{NPRG,KT}.
Moreover, the DS equations are given with
respect to the 
order parameter and, therefore, we can study 
only the symmetry broken phase.
Contrary to this, the NPRG enable us to 
describe the RG flows and the fixed points
irrespective of the phases.
Therefore, the RG approach is very suitable to study
of the conformal dynamics around the fixed point
as well as the phase structures \cite{KT}.
In addition, it will be found that the approximation
can be improved easily beyond the so-called ladder 
approximation in a somewhat systematic way in the
NPRG.
We will perform the RG analysis with an improved
framework from the conventional one later.
The phase structure of QCD with many flavors has been
also studied by the NPRG method \cite{Gies}.

Thanks to the simplicity of the NPRG equations,
we may easily extend the analysis to
gauge-Yukawa models with introducing a gauge
singlet scalar field.
Then it is found that the BZ fixed point becomes unstable
and an IR fixed point with a non-trivial Yukawa coupling
appears as a stable fixed point instead.
We will show the conformal dynamics around the new
fixed point and the phase diagram obtained by the
NPRG analysis.

As is well known,  the mass parameter of a scalar field
is relevant and fine-tuning is required in order
to adjust it near the critical surface, or to
make a large hierarchy in the mass scales.
However the scalar field acquires a large
anomalous dimension at the IR fixed point,
if the fixed point interaction is strong.
Then the anomalous dimension suppresses the mass 
parameter towards the IR direction and
the power of renormalization scale dependence may be
reduced to even almost logarithmic.
Therefore the fine-tuning aspect turns out to be
very different from that given in the weak coupling
region.

Recently, Luty and Okui \cite{LO} considered a
composite Higgs model, where the strong dynamics
is approximately conformal and the Higgs
field is endowed with a large anomalous dimension.
This model is given in the chiral symmetry broken
phase of a QCD like gauge theory and, however, 
is near the ultra-violet (UV) fixed point, 
which will be discussed later.
One of the author also proposed another type of the
Higgs model \cite{conformalhiggs}, 
which makes use of the IR fixed point 
with a non-trivial Yukawa coupling.
There the fine-tuning problem of the
standard model (SM) Higgs may be ameliorated owing to
the large anomalous dimension. 
Thus knowledge of the conformal dynamics of
gauge theories seems to be very useful even in
phenomenological studies, specially, of the
Higgs sector.

This paper is organized as follows.
We start in section~II with perturbative analysis
of the IR fixed point for the gauge-Yukawa models.
In section~III, we develop the NPRG equations for the
$SU(N_c)$ gauge theory with $N_f$ massless flavors.
Section~IV addresses the fixed points, anomalous 
dimensions as well as the phase diagrams of the
gauge theories by solving the NPRG equations.
In section~V, we extend the RG analysis to the 
gauge-Yukawa model and study the conformal dynamics
around the IR fixed point with a non-trivial Yukawa
coupling.
The effect of the large anomalous dimension of the
scalar field is also examined in section~VI
Finally section~VII is devoted to the conclusions and 
discussions.

%%%%%%%%%%%%%%%%%%%%%%%%%%%%%
\section{Perturbative analysis of the IR fixed points}

\subsection{Banks-Zaks (BZ) fixed point}

It is well-known that a IR fixed point, called
the BZ fixed point \cite{BZ}, 
realizes in an asymptotically
free gauge theory with appropriate numbers of flavors.
When we add a gauge singlet scalar field coupled 
through Yukawa interaction with the fermions,
then the BZ fixed point becomes unstable.
In some cases, the RG flows are found to approach 
a new fixed point with non-vanishing Yukawa coupling.
In this section, let us discuss the fixed points 
by using the perturbative RG equations.

We restrict our discussion to the $SU(N_c)$ gauge
theories with $N_f$ massless flavors of the $N_c$ 
dimensional representation.
The two loop beta function for the gauge
coupling $\alpha_g = g^2/(4\pi)^2$ is found to be
\be
\mu \frac{d \alpha_g}{d \mu}
= -2b_0 \alpha_g^2 - 2 b_1 \alpha_g^3 + \cdots,
\label{pgaugebeta}
\ee
where the coefficients are given with the
quadratic Casimir $C_2(N_c) = (N_c^2 -1)/2N_c$ as
\bea
b_0 &=& \frac{11}{3} N_c - \frac{2}{3}N_f,  \\
b_1 &=& \frac{34}{3} N_c^2 - \frac{N_f}{2}
\left(
4 C_2(N_c) + \frac{20}{3} N_c
\right).
\eea
The IR fixed point exists when $b_0 >0$ and $b_1 < 0$.
This constrains the conformal window for $N_f$ to be
$
(34/13)N_c < N_f < (11/2)N_c
$
in the large $N_c$ leading. 
The value of the fixed point gauge coupling,
$\alpha_g^* = - b_0/b_1$, increases as $N_f$ decreases.
As long as this fixed point coupling is sufficiently small,
or $N_f$ is close to $(11/2)N_c$,
the perturbative analysis is reliable.

However the lower bound of the conformal window is not
meaningful in practice, since the fixed point gauge coupling
is very strong with the flavor number near the lower bound.
Then such  perturbative calculation is not reliable.
Indeed, the three-loop beta function reduces
the gauge coupling at the fixed point considerably.
It does not mean that the three-loop result is
more reliable though, since 
the beta functions calculated by perturbation
give an asymptotic series.
Although it would be better to use the beta function
evaluated {\it e.g.} by means of the Borel summation
instead, it is beyond our present scope.
In this paper, we use the two-loop beta function simply
as the primitive step of the analysis.
Moreover, it is found that 
such strong gauge interaction 
induces spontaneous chiral symmetry breaking and
erases the fixed point.
It will be shown that the NPRG analysis combined with
the two-loop gauge beta function leads to 
the conformal window given by $4N_c < N_f < (11/2)N_c$,
which coincides with the result obtained 
by analyzing the ladder DS equations \cite{DS}.

In the following discussions, the anomalous dimension of 
the fermion mass operator $\bar{\psi}\psi$ plays an
important role. 
The anomalous dimension $\gamma_{\bar{\psi}\psi}$ evaluated 
at the one-loop level is $-6 C_2(N_c) \alpha_g$.
We note that this anomalous dimension is negative
at the BZ fixed point.
This implies that the Yukawa coupling with a gauge
singlet scalar is relevant there,
since the dimension of the operator
$\phi (\bar{\psi}\psi)$ is less than 4.
Thus it is seen that the BZ fixed point become
unstable under perturbation by the Yukawa interactions.

\subsection{Yukawa interaction with a ``meson'' field }

First let us consider to incorporate a Yukawa 
interaction respecting the flavor symmetry of 
the unperturbed theory.
The flavor symmetry of the gauge theory with 
$N_f$ massless flavors is
$U(N_f)_L \times U(N_f)_R$, although
the axial $U(1)$ part is broken by anomaly.
We introduce a ``meson''-like scalar $\Phi$,
which transforms as
$\Phi \rightarrow g_L \Phi g_R^{\dagger}$
with the group elements $g_{L(R)}$ of
$U(N_f)_{L(R)}$.
The invariant Yukawa interaction is given by
\be
{\cal L} \sim - y~\Phi_j^i \bar{\psi}_{Li} \psi_R^j
 + \mbox{h.c.}.
\label{mesonyukawa}
\ee
Then the beta functions for the gauge coupling
$\alpha_g = g^2/(4\pi)^2$ and the Yukawa coupling
$\alpha_y = |y|^2/(4\pi)^2$ are found to be \cite{MV}
\bea
\mu \frac{d \alpha_g}{d \mu}
&=& -2b_0 \alpha_g^2 - 2 b_1 \alpha_g^3 
- 2 N_f^2 \alpha_g^2 \alpha_y,  
\label{mesongaugebeta} \\
\mu \frac{d \alpha_y}{d \mu}
&=& 2 \alpha_y \left[
(2 N_c + N_f) \alpha_y - 6 C_2(N_c) \alpha_g
\right].
\label{mesonyukawabeta}
\eea
Here we evaluate the Yukawa beta function at the
one-loop order.

It is found from these beta functions that the 
IR fixed point exists only for 
$5.24 < N_f/N_c < 5.5$ in the large $N_c$ leading
and the fixed point gauge couplings
are very small in any case
\footnote{
The conformal window of 
$N=1$ supersymmetric gauge theories is known to be
$3/2 < N_f/N_c < 3$ \cite{duality}.
According to the Seiberg duality, the theories
in this window also have a fixed point with
a nontrivial Yukawa coupling, when a mesonic 
chiral superfield is introduced.
The RG flows and the fixed points may be explicitly
examined \cite{susyRG,KNT}.
The RG aspect of non-supersymmetric theories
is very different from the supersymmetric cases.}.
When the flavor number is less than this window,
then the RG flow goes infinity eventually.
This is because the non-trivial zero point
of the two-loop gauge beta function is eliminated
by a small Yukawa coupling.
The one-loop approximation of the Yukawa beta
coupling is not the origin.
Therefore if we use the three-loop beta function
or some non-perturbative beta function for the
gauge coupling, then the result may different.
This is beyond our present scope and we do not consider
this type of gauge-Yukawa models.

\subsection{Yukawa interaction with a singlet scalar}

Next we consider the Yukawa coupling with
a flavor singlet scalar $\phi$ such as
\be
{\cal L} \sim 
- \phi \sum_{i=1}^{n_f} y_i ~\bar{\psi}_{Li} \psi_R^i 
+ \mbox{h.c.},
\label{singletyukawa}
\ee
where $n_f = 1, \cdots, N_f$.
The Yukawa couplings can be diagonal zed by the
flavor rotation.
We note that these Yukawa couplings break the flavor
symmetry of $U(N_f)_L \times U(N_f)_R$ explicitly.

The RG equations for the gauge coupling and the Yukawa 
coupling are given explicitly by \cite{MV}
\bea
\mu \frac{d \alpha_g}{d \mu} 
&=&
-2 b_0 \alpha_g^2 - 2 b_1 \alpha_g^3 
- 2  \alpha_g^2 \sum_{i=1}^{n_f} 
\alpha_{y_i}, 
\label{singletgaugebeta}
\\
\mu \frac{d \alpha_{y_i}}{d \mu}
&=&
2 \alpha_{y_i} \left[
2 N_c \sum_{j=1}^{n_f} \alpha_{y_j} + \alpha_{y_i}
- 6 C_2(N_c) \alpha_g
\right],
\label{singletyukawabeta}
\eea
where we used the one-loop beta function for the Yukawa 
coupling again.
Extension of the Yukawa beta function to the two-loop
\cite{MV} order does not alter the fixed point 
structure significantly,
although 2-loop terms are essential for the gauge beta
function to see the BZ fixed point.

Let us examine the fixed point with non-trivial Yukawa 
couplings. We may set $y_i = y^*$, since
the Yukawa couplings are expected to be
the same at the IR fixed point.
Then the explicit beta functions,
(\ref{pgaugebeta}) and (\ref{singletyukawabeta}),
lead the non-trivial fixed point couplings 
$\alpha_g^*$ and $\alpha_y^*$ given by
\bea
& &
\alpha_g^* = \frac{b_0}{-b_1 + \frac{6C_2(N_c) n_f}{2N_c n_f + 1}}
\simeq \frac{b_0}{-b_1 + 3/2}, 
\label{pgaugefp}\\
& &
\alpha_y^* = \frac{6C_2(N_c)}{2 N_c n_f + 1} \alpha_g^*
\simeq \frac{3}{2n_f}\alpha_g^*,
\label{pyukawafp}
\eea
in the large $N_c$ leading.
Thus the new IR fixed point is found to always appear, 
since $b_1 < 0$ for existence of the BZ fixed point.
We will examine the dynamics in this type of
gauge-Yukawa models by means of
the NPRG in section~V and section~VI.

\section{Non-perturbative RG for gauge theories}

\subsection{Approximation scheme and NPRG equations}

In practice it is a quite difficult problem to extend the 
RG equations to ones fully reliable even in the 
non-perturbative region.
Here we consider to apply the so-called exact renormalization
group (ERG) \cite{ERG}
reduced by some approximation scheme,
which we call the NPRG.
It has been found \cite{NPRG,KT} that the 
NPRG method can also describe
the spontaneous chiral symmetry breaking
phenomena induced by strong gauge interaction
and reproduces the results obtained by solving
the DS equations.
We first review the formulation briefly.

The ERG equation gives evolution of the Wilsonian
effective action under infinitesimal shift of the cutoff
scale.
%There have been several ERG equations, the Wegner-Houghton
%equation \cite{WH}, the Polchinski equation \cite{polchinski}, 
%the Legendre flow equation \cite{legendre} and so on.
Although it is amusing that the evolution can be represented
exactly as a functional equation with respect to the
Wilsonian effective action,
it is necessary to reduce the equations by some
approximation in order to analyze them.
It is usually performed to truncate the series of
local operators in the Wilsonian 
effective action.
Improvement of the approximation is made by 
increasing the level of the operator truncation.

Once the operator truncation is performed,
then the ERG equation turns out to be a set of
one-loop RG equations.
Difference from the perturbative RG lies in the point
that couplings of the higher dimensional operators
are involved as well as the renormalizable operators.
This procedure enables us to sum up an infinite
number of loop diagrams.

It was found through the previous studies \cite{NPRG} that 
the effective four-Fermi operators are found to 
play an important role for the non-perturbative 
analysis of the chiral symmetry breaking.
The reason may be understood by thinking over
the anomalous dimension of the fermion
mass operator $\gamma_{\bar{\psi}\psi}$.
Fig.~1 shows schematically 
how the anomalous dimension 
is represented in terms of the effective four-Fermi
couplings in the NPRG framework.
The four-Fermi couplings are also given as a sum of
infinitely many loop diagrams by solving the RG
equations.
Thus a non-perturbative sum of the loop diagrams
is carried out by incorporating the four-Fermi 
operators.

\begin{figure}[htb]
\includegraphics[width=0.5\textwidth]{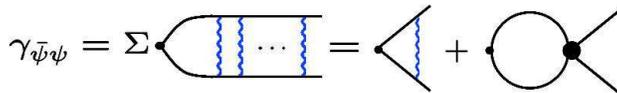}
\caption{\label{fig1} 
The anomalous dimension $\gamma_{\bar{\psi}\psi}$
in terms of the effective four-Fermi vertex is
shown schematically.
}
\end{figure}

Here we shall consider the four-Fermi operators induced
in the Wilsonian effective action of the gauge theory.
The four-Fermi operators should be invariant under the
color $SU(N_c)$ as well as the
flavor symmetry $U(N_f)_L \times U(N_f)_R$.
Among many invariant operators,
we take only the following operators
\footnote{
The effective four-fermi operators in QCD with many flavors
and their renormalization have been also discussed
in Refs.~\cite{Gies}.
The RG equations given there may be related with
our results.
},
\bea
{\cal O}_S & = &
2 \bar{\psi}_{Li} \psi_R^j ~\bar{\psi}_{Rj} \psi_L^i,  
\label{OS}\\
{\cal O}_V & = &
\bar{\psi}_{Li} \gamma_{\mu} \psi_L^j~
\bar{\psi}_{Lj} \gamma_{\mu} \psi_L^i
+ (L \leftrightarrow R),  
\label{OV}\\
{\cal O}_{V1} & = &
2 \bar{\psi}_{Li} \gamma_{\mu} \psi_L^i~
\bar{\psi}_{Rj} \gamma_{\mu} \psi_R^j,  
\label{OV1}\\
{\cal O}_{V2} & = &
(\bar{\psi}_{Li} \gamma_{\mu} \psi_L^i)^2
+ (L \leftrightarrow R).
\label{OV2}
\eea
Then the four-fermi part of the Wilsonian
effective Lagrangian is given by
\be
-{\cal L_{\rm eff}} \sim
\frac{G_S(\Lambda)}{\Lambda^2} {\cal O}_S
+ \frac{G_V(\Lambda)}{\Lambda^2} {\cal O}_V
+ \frac{G_{V1}(\Lambda)}{\Lambda^2} {\cal O}_{V1}
+ \frac{G_{V2}(\Lambda)}{\Lambda^2} {\cal O}_{V2},
\label{4fermilagrangian}
\ee
where the couplings $G$ are dimensionless. 
The operators which contain
the $SU(N_c)$ generators $T^A$ may be
reduced to the above operators by performing the
Fierz transformation; {\it e.g.}
\bea
&&\sum_A
\bar{\psi}_{Li} T^A \gamma_{\mu}\psi_L^i~
\bar{\psi}_{Rj} T^A \gamma_{\mu}\psi_R^j
= - \frac{C_2(N_c)}{N_c}
{\cal O}_S + \cdots,  \\
&&\sum_A
\left[
\left(
\bar{\psi}_{Li} T^A \gamma_{\mu}\psi_L^i
\right)^2
+
(L \leftrightarrow R)
\right]
= \frac{C_2(N_c)}{N_c}
{\cal O}_V + \cdots.
\eea
We also discard the higher dimensional
operators containing covariant derivatives
in the lowest order of truncation. 
It is noted that the four-fermi operators are not
induced through any local operators with 8 fermions
or more.
Therefore the RG flows of the four fermi couplings
are not influenced by the higher dimensional operators
as long as we take only the local operators.

We should also incorporate higher dimensional
operators of the field strength, such as
$(D_{\mu} F^{\mu \nu})^2$, 
$(F_{\mu \nu}F^{\mu \nu})^2$ and so on.
Otherwise the higher loop part of the gauge beta
function in Eq.~(\ref{pgaugebeta}) is not reproduced.
However practical calculations of the loop diagrams
are rather tedious.
It is also a hard problem to maintain the gauge invariance
in the Wilson RG, since the RG is defined with 
cutoff the momentum scale. For the recent
approach to this problem, see Refs.~\cite{Morris}.
In addition, it would be enough to follow the RG flow of 
the gauge coupling only for the present purpose.
Therefore we do not deal with the ERG equations
faithfully, but substitute the two-loop beta
function (\ref{pgaugebeta}) 
for the RG equation of the gauge coupling instead.
The approximation scheme of the present
analysis is given by these procedures.

\subsection{The RG equations}

Now it is enough to deduce the RG equations for the
four-fermi couplings $G_S$, $G_V$, $G_{V1}$ and
$G_{V2}$ in the simplified scheme explained above.
Calculation of the RG equations is rather lengthy but 
straightforward.
In Appendix, we consider the one-loop corrections
by the gauge interaction explicitly and illustrate
how we may calculate the RG equations with these examples.
Then it is found that only the four-fermi couplings
$G_S$ and $G_V$ are induced.
It is also shown that the gauge corrections are dependent
on the gauge choice.
Here we take the Landau gauge, in which the wave
function renormalization of the fermions vanishes.

There are also other loop diagrams incorporating two
four-fermi operators and so on.
The resultant RG equations are shown in Appendix.
In the followings analysis, however, we extract
the large $N_c$ and $N_f$ leading part from them
for the simplicity.
Then it is seen that the equations for $G_S$ and $G_V$
decouple from $G_{V1}$ and $G_{V2}$.
Besides the couplings $G_{V1}$ and $G_{V2}$ are not
induced at all, if their initial values are vanishing.
Thus it is all right to examine the RG equations for
$G_S$ and $G_V$, which are given explicitly by
\bea
\Lambda \frac{d g_S}{d \Lambda}
&=& 2g_S
- 2 N_c \left(
g_S + \frac{3}{2} \alpha_g
\right)^2 + 2 N_f g_S g_V,  
\label{largeNGSRG}
\\
\Lambda \frac{d g_V}{d \Lambda}
&=& 2 g_V 
+(N_c + N_f) g_V^2 + \frac{1}{4}N_f g_S^2
-\frac{3}{4}N_c \alpha_g^2,
\label{largeNGVRG}
\eea
where we rescaled the couplings as
$g_{S(V)} = G_{S(V)}/(4\pi^2)$.
We solve the RG equations
(\ref{largeNGSRG}) and (\ref{largeNGVRG})
coupled with the two-loop gauge beta function 
in the large $N_c$ and $N_f$ leading,
\be
\Lambda \frac{d \alpha_g}{d \Lambda}
= -2 N_c \left(
\frac{11}{3} - \frac{2N_f}{3N_c}
\right) \alpha_g^2
-2 N_c^2 \left(
\frac{34}{3} - \frac{13N_f}{3N_c}
\right)\alpha_g^3.
\ee

On the other hand, solutions of the DS equations
are usually given by sum of the ladder diagrams as
shown in Fig.~1. 
Therefore most of the analyses using the SD equations
have been carried out in the ladder approximation. 
We may restrict the loop corrections taken in the RG 
equations (\ref{largeNGSRG}) and (\ref{largeNGVRG})
to the ladder diagrams as well.
Then the last term in Eq.~(\ref{largeNGSRG}) is
found to drop out and the RG equation is reduced to
\be
\Lambda \frac{d g_S}{d \Lambda}
= 2g_S
- 2 N_c \left(
g_S + \frac{3}{2} \alpha_g
\right)^2.
\label{ladderRG}
\ee
It is noted that the coupling $g_V$ is not involved
with evolution of $g_S$ any more.
However the flavor number $N_f$ must be fairly large in
the cases involving the BZ fixed point.
Therefore the last term in Eq.~(\ref{largeNGSRG})
is supposed to give a sizeable effect.
In the next section, we also solve Eq.~(\ref{ladderRG})
in order to compare the results with those
obtained in the ladder approximation.

\section{Fixed points and anomalous dimensions}

The two-loop gauge beta function (\ref{pgaugebeta}) gives
the fixed point coupling as
\be
N_c \alpha_g^* =
\frac{11 - 2r}{13r-34},
\ee
where $34/13 < r=N_f/N_c < 11/2$.
However, this condition for the
flavor number does not always guarantee
existence of the IR fixed point.
All the other couplings must approach their
fixed point values.
The fixed point for the four-fermi coupling
$g_S$ in the ladder approximation can be
found analytically from the RG equation 
(\ref{ladderRG}).
Those are given by
\be
N_c g_S^* =
\frac{1}{2}\left(
1 - 3 N_c \alpha_g^* \pm 
\sqrt{1 - 6 N_c \alpha_g^*}
\right),
\label{ladderfp}
\ee
where the solution with $+(-)$ gives the UV(IR) 
fixed point.

It is noted that the fixed point does not exist for 
$N_c \alpha_g^* > 1/6$ and the four-fermi coupling
$g_S$ of any RG flow goes to infinity. 
This behavior of the RG flows shows that there 
is only broken phase of the chiral symmetry \cite{NPRG}.
Then the conformal window is found to be
\be
 4N_c \leq N_f \leq \frac{11}{2}N_c,
\ee
in the large $N_c$ leading.
However we should keep in mind that the lower bound
of this window is not robust owing to large dependence
on the gauge beta function, as was mentioned in 
section~II.

It is easy to find the fixed points in the non-ladder
approximation numerically from Eqs.~(\ref{largeNGSRG}) 
and (\ref{largeNGVRG}).
Then the fixed point couplings are given by 
$(\alpha_g^*, g_S^*, g_V^*)$.
In Fig.~2, the couplings $N_c (\alpha_g^*, g_S^*)$
of the IR (BZ) fixed points and the UV fixed points
are shown by taking  $r=N_f/N_c$ to be a 
continuous parameter. 
For the comparison, the
fixed points obtained from Eq.~(\ref{ladderfp})
in the ladder approximation are also shown by 
dotted line.

\begin{figure}[htb]
\includegraphics[width=0.5\textwidth]{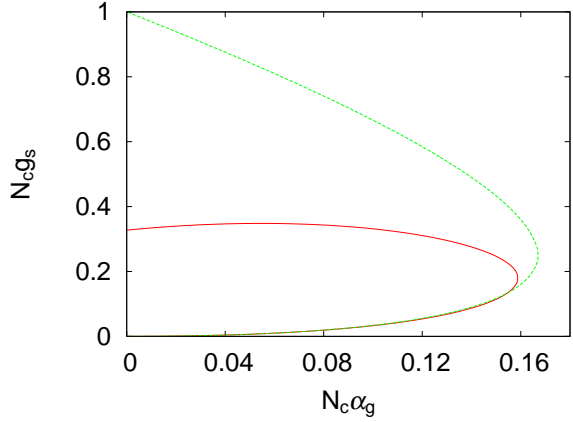}
\caption{\label{fig2}
The couplings $N_c (\alpha_g^*, g_S^*)$
of the IR (BZ) fixed points and the UV fixed points
are shown by taking  $r=N_f/N_c$ to be a 
continuous parameter. 
The dotted line represent the
fixed points obtained in the ladder approximation.
}
\end{figure}

We may figure out the RG flows in the coupling
space of $(\alpha_g, g_S, g_V)$ by solving the
RG equations numerically.
Fig.~3 shows some RG flows approaching the BZ fixed
point, which is represented by the point A, 
in the case of $r = N_f/N_c = 4.1$ for an 
illustration.
The points B and C stand for the UV fixed points.
It is seen how these fixed points are linked by
the renormalized trajectories. 
The renormalized trajectory connecting the fixed
points B and C lies in the phase boundary.
Some of the flows go along the critical surface
and approach the renormalized trajectory.

In Fig.~4 the RG flows in the cross section 
of $\alpha_g = \alpha_g^*$ are also shown.
The critical surface in the $(g_S, g_V)$
space as well as the renormalized trajectory
connecting the fixed points A and B are
clearly seen.
In the lower phase, where all RG flows approach the
IR fixed point, the IR theory appears as
a CFT. 
On the other hand, the chiral symmetry is spontaneously
broken in the upper phase, where the four-fermi 
coupling $g_S$ goes to infinity.

\begin{figure}[htb]
\includegraphics[width=0.6\textwidth]{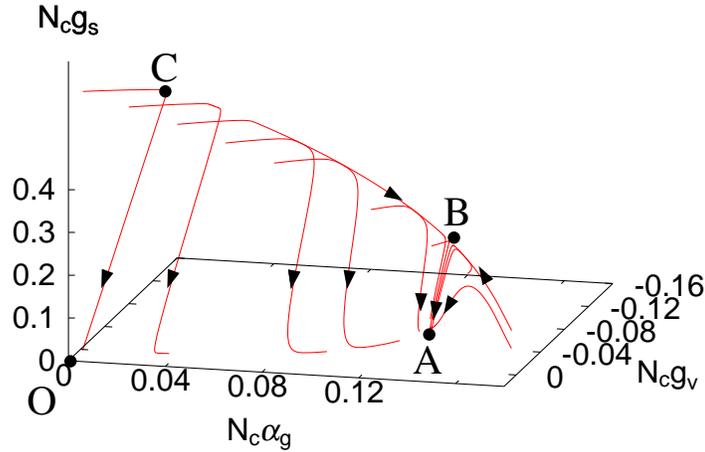}
\caption{\label{fig3}
The fixed points and some RG flows towards the IR
direction are shown in the coupling space of 
$N_c (\alpha_g, g_S, g_V)$ for 
$r = N_f/N_c = 4.1$ as an example. 
The point A represents the BZ fixed point.
The points B and C stand for UV fixed points
and the line connecting them lies in the 
phase boundary.
}
\end{figure}

\begin{figure}[htb]
\includegraphics[width=0.5\textwidth]{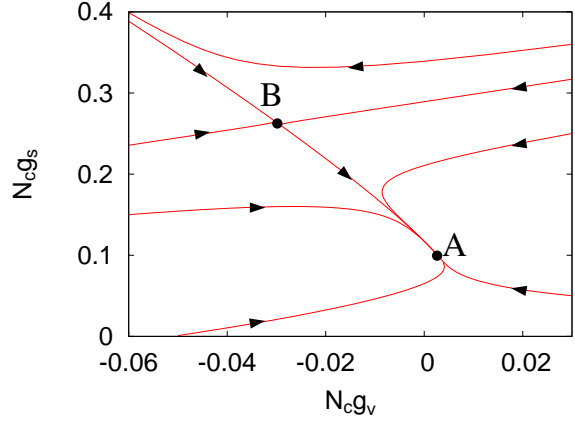}
\caption{\label{fig4}
The RG flows and the fixed points A and B
in the cross section of $\alpha_g = \alpha_g^*$
are shown for $r = N_f/N_c = 4.1$.
}
\end{figure}

The phase divides the cutoff gauge theories given with
various gauge couplings into two classes.
The critical value of the gauge coupling 
$\alpha_g^{\rm cr}$ can be
read off from the phase diagram.
In Fig.~5, $N_c \alpha_g^{\rm cr}$
is shown for various $r=N_f/N_c$.
The dotted line stands for the critical
gauge coupling obtained in the ladder approximation. 

\begin{figure}[htb]
\includegraphics[width=0.5\textwidth]{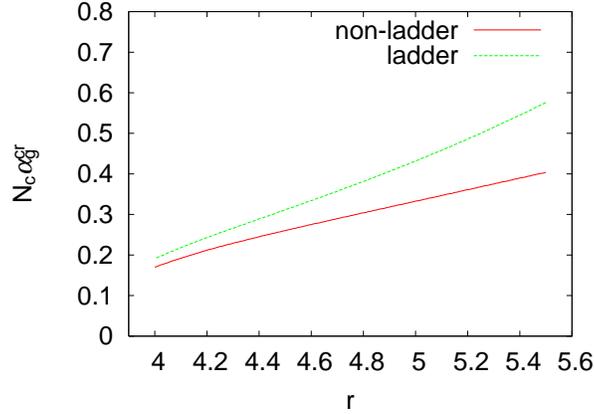}
\caption{\label{fig5}
The critical gauge coupling $N_c \alpha_g^{\rm cr}$
is given for various $r=N_f/N_c$.
The dotted line stands for the same obtained 
in the ladder approximation. 
}
\end{figure}

Next we discuss the anomalous dimension of the
four-fermi operators at the fixed points A and B.
It should be also stressed that the NPRG offers
us the right method to evaluate the anomalous 
dimensions, since they are obtained directly from the
RG equations.
First we shall consider the anomalous dimensions of
the four-fermi operators, which may be easily
evaluated by taking
infinitesimal deviation of the four-fermi couplings
from the fixed point,
$g_S = g_S^* + \delta g_S$ and
$g_V = g_V^* + \delta g_V$.
Then the deviations follow the RG equations,
\be
\Lambda \frac{d }{d \Lambda}
\left(
\begin{array}{c}
\delta g_S \\
\delta g_V
\end{array}
\right)
= 
\left(
\begin{array}{cc}
2 - 4 N_c g_S^* - 6N_c \alpha_g^* + 2 N_f g_V^* & 2 N_f g_S^* \\
(1/2) N_f g_S^* & 2 + (N_c + N_f)g_V^*
\end{array}
\right)
\left(
\begin{array}{c}
\delta g_S \\
\delta g_V
\end{array}
\right).
\ee
The eigenvalues of the matrix in the right hand side,
which we write by $\lambda_1$ and $\lambda_2$ 
$(\lambda_1 < \lambda_2)$,
represent nothing but the scaling dimensions of 
the four-fermi couplings.
The four-fermi operators for the eigenmodes, 
${\cal O}_1$ and ${\cal O}_2$, 
are given by linear combinations
of ${\cal O}_S$ and ${\cal O}_V$.
Then the anomalous dimensions of these operators are
given by $\gamma_{{\cal O}_i}^* = \lambda_i - 2$
$(i=1,2)$.

It is found that the both eigenvalues are positive
at the IR fixed point A.
This means that there is no relevant operators
at the fixed point.
On the other hand, $\lambda_1$ turns out to be negative
at the UV fixed point B.
This relevant direction gives the renormalized trajectory
towards the IR fixed point.
In Fig.~6, the anomalous dimensions
$|\gamma_{{\cal O}_1}^*|$ are shown with respect to
the fixed point gauge couplings $N_c \alpha_g^*$.

In the ladder approximation, the anomalous dimension
of the four-fermi operator ${\cal O}_S$ may
be evaluated analytically.
It is found from the RG equation (\ref{ladderRG}) 
immediately as
\be
\gamma_{{\cal O}_S}^* 
= -4N_c  g_S^* - 6N_c \alpha_g^*
= -2 \left( 
1 \pm \sqrt{1 - 6 N_c \alpha_g^*}
\right).
\ee
This should be compared with $\gamma_{{\cal O}_1}^*$
evaluated in the non-ladder calculation.
The dashed line in Fig.~6 stands for 
$|\gamma_{{\cal O}_S}^*|$ in the ladder approximation.
It is seen that the anomalous dimension of the
relevant four-fermi operator becomes slightly larger
by incorporating the non-ladder diagrams, but
the difference is not significant.

\begin{figure}[htb]
\includegraphics[width=0.5\textwidth]{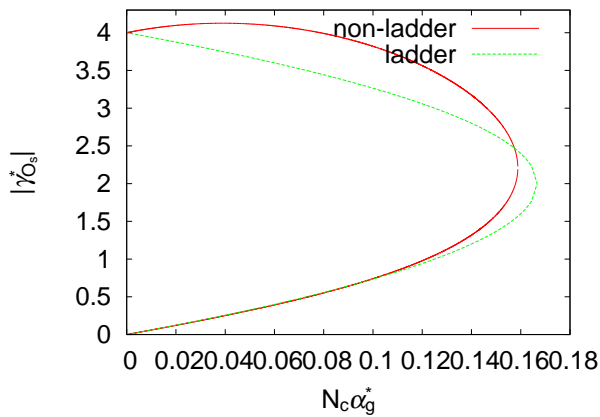}
\caption{\label{fig6}
The anomalous dimensions $|\gamma_{{\cal O}_1}^*|$
at the fixed points with respect to $\alpha_g^*$
for various flavor numbers are shown.
The dashed line stands for 
$|\gamma_{{\cal O}_S}^*|$ obtained in the
ladder approximation.
}
\end{figure}

Another important quantity in our discussion
is the anomalous dimension of the fermion mass 
operator $\bar{\psi}\psi$.
It may be evaluated by introducing
an infinitesimally small mass to the vector-like
fermions.
The anomalous dimension is found to be
\be
\gamma_{\bar{\psi}\psi}
= - 6 C_2(N_c)\alpha_g - 2 N_c g_S + 8 g_{V1}
\simeq - N_c (3\alpha_g + 2  g_S),
\ee
in the large $N_c$ leading.
In the ladder approximation, it is easily 
obtained as
\be
\gamma_{\bar{\psi}\psi}^* 
= \frac{1}{2}\gamma_{(\bar{\psi}\psi)^2}^*
= 1 \pm \sqrt{1 - 6 N_c \alpha_g^*}.
\label{gammafermimass}
\ee
This result at the UV fixed point has been
also found by solving the ladder DS
equation \cite{DSanomalous}.
Fig.~7 shows the anomalous dimension  
$|\gamma_{\bar{\psi}\psi}^*|$ with respect to the
fixed point gauge coupling
$N_c \alpha_g^*$ for various flavor numbers.
The dashed line stands for the same in the
ladder approximation.
Contrary to the anomalous dimensions of the
four-fermi operator, $|\gamma_{\bar{\psi}\psi}^*|$
obtained in the non-ladder approximation are rather
different from Eq.~(\ref{gammafermimass}).
However, the anomalous dimensions  at the IR fixed point
are fairly stable under improvement
of the approximation.
Therefore, the results may be thought to be
reliable even in the strong coupling region.
This is a good point of our analysis, 
since the explicit values of the
anomalous dimensions $|\gamma_{\bar{\psi}\psi}^*|$
will be important in the following observations.

\begin{figure}[htb]
\includegraphics[width=0.5\textwidth]{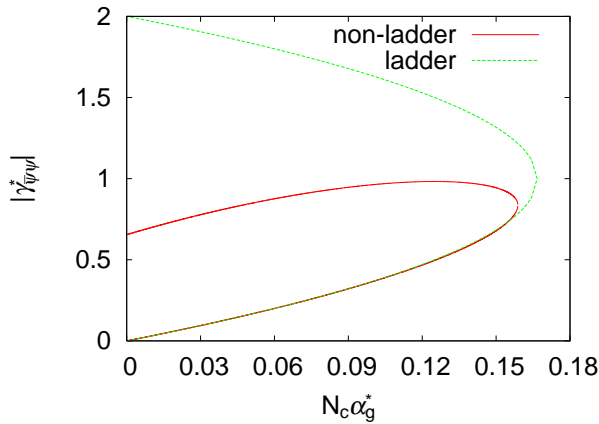}
\caption{\label{fig7}
The anomalous dimensions $|\gamma_{\bar{\psi}\psi}^*|$
at the fixed points.
The dashed line stands for the same in the
ladder approximation.
}
\end{figure}

\section{IR fixed point in the gauge-Yukawa models}

Now we extend the previous analysis
to the gauge-Yukawa models.
We consider the Yukawa interaction with a flavor
singlet scalar $\phi$ given by Eq.~(\ref{singletyukawa}).
In order to examine dynamics around the IR
fixed point, let us set all Yukawa coupling
to be equal, $y_i = y$ $(i=1,\cdots, n_f)$.

The flavor symmetry is explicitly broken by
the Yukawa interaction.
Therefore the operators,
which are not invariant under
$U(N_f)_L \times U(N_f)_R$, such as
$\bar{\psi}_{Li} \psi_R^i~
\bar{\psi}_{Rj} \psi_L^j$
and
$\bar{\psi}_{Li} \gamma_{\mu} \psi_L^j~
\bar{\psi}_{Rj} \gamma_{\mu} \psi_R^i$
are induced by loop corrections
However the diagrams with scalar lines may be 
just omitted as sub leading contributions
in the large $N_c$ and $N_f$ leading. 
Therefore we do not have to incorporate
other four-fermi operators than ${\cal O}_S$
and ${\cal O}_V$ in this limit.
Besides, we may apply the same beta functions 
for the four-fermi couplings as well as for 
the gauge beta function used in the previous
section.
Only the RG equation of the Yukawa coupling
should be modified. 

The RG equation of the Yukawa coupling 
$\alpha_y = |y|^2/(4\pi)^2$ is given
in terms of the anomalous dimensions of the
scalar field $\gamma_{\phi}$ and 
$\gamma_{\bar{\psi}\psi}$ as
\be
\Lambda\frac{d \alpha_y}{d \Lambda} = 
2 \alpha_y \left(
\gamma_{\phi} + \gamma_{\bar{\psi}\psi}
\right).
\ee
Here it is noted that  the four-fermi operators
are also involved in the anomalous dimension
$\gamma_{\bar{\psi}\psi}$ in the NPRG framework.
Explicitly the anomalous dimensions are given by
\bea
\gamma_{\phi} &=& 2 N_c n_f \alpha_y, \\
\gamma_{\bar{\psi}\psi} &=&
\alpha_y - 6 C_2(N_c) \alpha_g -2 N_c g_S + 8g_{V1}.
\eea
Therefore the beta function for the Yukawa coupling 
is found to be 
\be
\Lambda\frac{d \alpha_y}{d \Lambda}
\simeq
2 \alpha_y \left[
2 N_c n_f \alpha_y - 3N_c \alpha_g -2 N_c g_S
\right],
\ee
in the large $N_c$ leading.

At the IR fixed point, if it exists, the beta
function of the Yukawa coupling should vanish.
Therefore the anomalous dimension $\gamma_{\phi}$
and the value of the Yukawa coupling
$\alpha_y^*$ at the fixed point are simply given by
\be
\gamma_{\phi}^* = 2 n_f N_c \alpha_y^* =
- \gamma_{\bar{\psi}\psi}^*,
\ee
which is always positive.
In this simple approximation scheme, 
the beta functions for $\alpha_g, g_S, g_V$
are independent of the Yukawa coupling.
Therefore the IR fixed
point with a non-trivial Yukawa coupling exists
in any case in the conformal window.
The anomalous dimension $\gamma_{\bar{\psi}\psi}^*$
as well as the fixed point couplings
of $(\alpha_g^*, g_S^*, g_V^*)$
are the same as those given at the BZ fixed point.
Therefore the lines shown in Fig.~7 also
represent the anomalous dimension of the scalar field
$\gamma_{\phi}^*$ at the IR fixed points.

In Fig.~8, the RG flows and the fixed points
are described in the coupling space of
$N_c(\alpha_y, \alpha_g, g_S)$ in the case of
$N_f = 4.1 N_c$ and $n_f=1$.
These flows are obtained by solving the RG equations
in the ladder approximation for simplicity.
The point A' stands for the IR fixed point of the
gauge-Yukawa theory, while the point A represents
the BZ fixed point.
The fixed points, A, B, B' are all unstable towards
the IR direction and are linked by the renormalized
trajectories as shown in Fig.~8.

\begin{figure}[htb]
\includegraphics[width=0.6\textwidth]{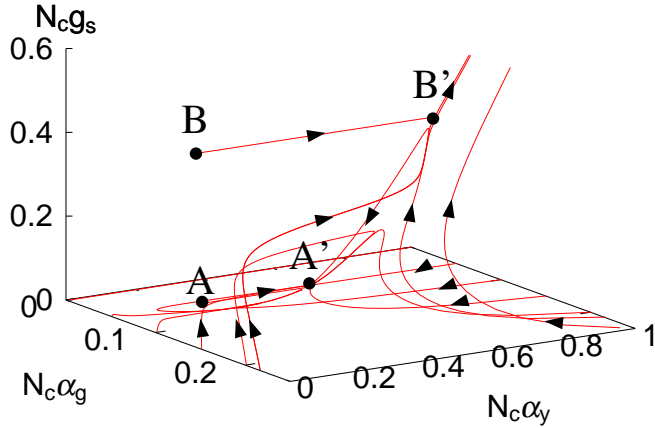}
\caption{\label{fig8}
The RG flows are described in the case of $N_f/N_c=4.1$
and $n_f=1$. The blob points A, A', B, B'
represent the fixed points.
The point A' is the IR fixed point, while
A is the BZ fixed point.
}
\end{figure}

Fig.~9 also describes the RG flows in the same case
so that the critical surface of the gauge-Yukawa
theory is seen clearly.
All the flows correspond to theories near critical
in the unbroken phase.
The phase structure of the gauge-Yukawa theories
has been also studied by means of the DS method 
\cite{KSTY,Hosek}.
In Ref.~\cite{KSTY}, the RG flows in the
broken phase are also given, but with fixed
gauge coupling constants.
The points C and C' represent the UV fixed point
of the pure four-fermi theories and of the four-fermi
theories with Yukawa interaction.
It is also found how the fixed points are linked with
the renormalized trajectories.

\begin{figure}[htb]
\includegraphics[width=0.6\textwidth]{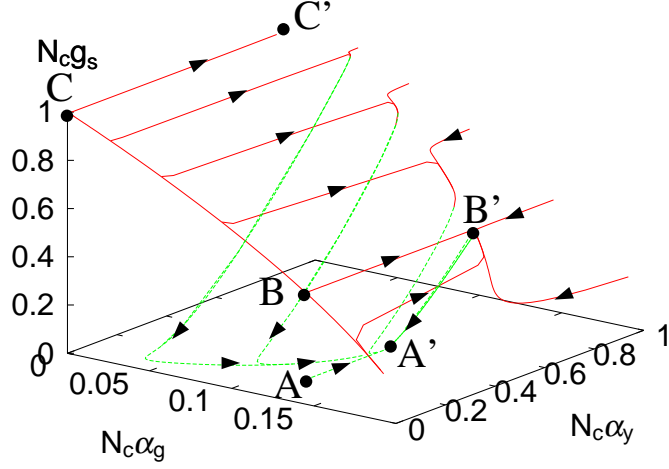}
\caption{\label{fig9}
The RG flows running in the very vicinity of the 
critical surface are shown in the same case with
Fig.~8.
The points C and C' represent the UV fixed point
of the pure four-fermi theories and of the four-fermi
theories with Yukawa interaction.
}
\end{figure}

When $N_f$ is equal to $4N_c$, then the UV fixed point
B' (B) and the IR fixed point A' (A) coincide with
each other and the relevant
four-fermi operator becomes exactly marginal at the
fixed point.
In practice, however, the non-leading contributions
with respect to $N_c$ and $N_f$ expansion cannot
be neglected. 
Let us also examine an explicit case of $N_c=3$ and
$N_f=12$ as an example.
Then the gauge coupling of the BZ fixed point
is given by $\alpha_g^* = 0.06$, which
is slightly less than the maximal coupling
$\alpha^*_{g, {\rm max}} = 1/12C_2(N_c) = 1/16$ 
in the conformal window \cite{ATW,MY}.
{}For the gauge-Yukawa theories, 
we apply the gauge beta function given by (8), which
depends on the Yukawa coupling at the two-loop level.
The beta function for the Yukawa coupling
is given by (30), (31) and (32).

As for the four-fermi couplings, 
let us restrict them to $g_S$ and $g_V$ only
for simplicity.
Then the beta functions for these couplings are
found to be
\bea
\Lambda\frac{d g_S}{d \Lambda}
&=& (2 + 2 \alpha_y) g_S
- 2 N_c g_S^2 + 2 N_f g_S g_V 
-12 C_2(N_c) \alpha_g g_S
+ 4 \alpha_y g_V  
\label{gsbeta} \\
& &
- 18 \frac{C_2(N_c)^2}{N_c} \alpha_g^2
-6\frac{C_2(N_c)}{N_c}
\left(
C_2(N_c) - \frac{N_c}{2}
\right)\alpha_g^2, \\
\Lambda\frac{d g_V}{d \Lambda}
&=& (2 + 2 \alpha_y) g_V
+ \frac{1}{4} N_f g_S^2
+ (N_c + N_f)g_V^2 +  \alpha_y g_S  \nn \\
& &
- 3 \frac{C_2(N_c)^2}{N_c} \alpha_g^2
-9\frac{C_2(N_c)}{N_c}
\left(
C_2(N_c) - \frac{N_c}{2}
\right)\alpha_g^2,
\label{gvbeta}
\eea
where note that the anomalous dimension of the fermion
field $\gamma_{\psi}$ is not vanishing but is given
by the Yukawa coupling in the Landau gauge.
In practice, we may find the IR fixed point and
the anomalous dimension $\gamma_{\phi}^*$ turns
out to be 0.637, when $n_f=1$.
If we take only the ladder diagrams, then the RG equation
is reduced to
\be
\Lambda\frac{d g_S}{d \Lambda}
= (2 + 2 \alpha_y) g_S
- 2 N_c \left(
g_S + \frac{3C_2(N_c)}{N_c}\alpha_g
\right)^2.
\label{ladderbeta2}
\ee
Then we may easily evaluate the anomalous dimension at the
IR fixed point as $\gamma_{\phi}^* = 0.65, 0.79, 0.95$ for 
$n_f = 1, 3, 12$ respectively.
Thus fairly large anomalous dimensions may be realized
in the gauge-Yukawa theories with the critical flavor
number $N_f = 4 N_c$. 

\section{Renormalization of the scalar potential}

Renormalization of the scalar potential is
peculiar, when the anomalous dimension is large.
In this section, we discuss renormalization
of the scalar mass $m^2_{\phi}$ and the quartic
interaction coupling $\lambda_4$.
When we expand the scalar potential 
at the scale $\Lambda$ as
\be
V(\phi) =
\tilde{m}_{\phi}^2 \Lambda^2
|\phi|^2 + \frac{\lambda_4}{4} |\phi|^4
+ \cdots,
\ee
then 
the RG equations for the dimensionless 
parameters $\tilde{m}_{\phi}^2$ and $\lambda_4$
are found to be
\bea
\Lambda \frac{d \tilde{m}^2_{\phi}}{d \Lambda} &=&
-2(1 - \gamma_{\phi}) \tilde{m}^2_{\phi} + 
4N_c n_f \alpha_y - 2 \tilde{\lambda}_4, 
\label{scalarmassRG}\\
\Lambda \frac{d \tilde{\lambda}_4}{d \Lambda} &=&
4 \gamma_{\phi}\tilde{\lambda}_4 
- 8 N_c n_f \alpha_y^2 + \tilde{\lambda}_4^2,
\label{quarticRG}
\eea
where $\tilde{\lambda}_4 = \lambda_4/(4\pi)^2$ and
$\gamma_{\phi} = 2 N_c n_f \alpha_y$.

First we shall discuss analytical solutions of
these equations, when the
Yukawa coupling stays on the IR fixed point 
$\alpha_y = \alpha_y^*$.
We also take the large $N_c$ leading.
Then contributions by the quartic coupling
$\tilde{\lambda}_4$ in (\ref{scalarmassRG}) 
and (\ref{quarticRG}) may be ignored. 
Note that the anomalous dimension $\gamma_{\phi}^*$
at the fixed point can
be as large as 1 in the strongly coupled case, 
as was seen in the previous section.
Therefore the effective dimension of the
scalar mass $m^2_{\phi}$, which is given by
$2(1 - \gamma_{\phi}^*)$, 
can be rather small, as $N_f$ approach the critical value.
However, $m^2_{\phi}$ is always relevant,
since  $\gamma_{\phi}^*$ is less than 1.

The solution of the scalar mass at a lower scale $\mu$
may be written down as 
\be
m_{\phi}^2(\mu) 
=
\left[
\tilde{m}_{\phi}^{*2} + \left(
\frac{\Lambda}{\mu}\right)^{\epsilon}
\left(
\tilde{m}_{\phi}^2(\Lambda) - \tilde{m}_{\phi}^{*2}
\right)
\right]\mu^2,
\label{masssol}
\ee
where $\epsilon \equiv 2(1-\gamma_{\phi}^*)$ and
$\tilde{m}_{\phi}^{*2} \equiv 2 \gamma_{\phi}^*/\epsilon
= \gamma_{\phi}^*/(1 - \gamma_{\phi}^*)$.
The power of divergence is reduced to $\epsilon$
due to the anomalous dimension.
As the renormalization scale $\mu$ goes down to zero,
$m_{\phi}^2(\mu)$ also decreases to zero
irrespectively of the initial value 
$\tilde{m}_{\phi}^2(\Lambda)$.
Thus  the theory on the IR fixed point describes
a massless interacting scalar.
It should be also noted that $\tilde{m}_{\phi}^{*2}$
gives the fixed point value of the dimensionless scalar 
mass parameter.
When $\tilde{m}_{\phi}^2$ is set to this value,
the whole theory is scale invariant in the quantum
mechanical sense.

Here let us consider the fine-tuning required
in order to realize a specific scalar mass
$m_{\phi}^2(\mu)= A \mu^2$ at a low energy 
scale $\mu$, where $A$ is a parameter of O(1).
The degree of fine-tuning may be given by
\cite{BG}
\be
C=\left|
\frac{\delta m^2_{\phi}(\mu)}{\delta m^2_{\phi}(\Lambda)}
\frac{m^2_{\phi}(\Lambda)}{m^2_{\phi}(\mu)}
\right|.
\label{finetuning}
\ee
Then this degree may be evaluated by using the explicit
solution (\ref{masssol}) and is found to be
\be
C = \frac{\tilde{m}_{\phi}^{*2}}{A}
\left(
\frac{\Lambda}{\mu}\right)^{\epsilon}
+ (A - \tilde{m}_{\phi}^{*2}).
\ee
Here the second term may be neglected, since
$\Lambda/\mu$ is much larger than 1 when we are 
concerned about fine-tuning.
The degree of fine-tuning $C$ increases, as the
cutoff scale $\Lambda$ is enlarged.
Therefore there is maximal value of $\Lambda$
so that the fine-tuning is remained less than $C$,
and it may be evaluated as
$\Lambda = O(1) \times C^{1/\epsilon} \mu $.
When $\epsilon=0.4$, this cutoff scale is 100 times
enlarged compared with the case of $\epsilon=2.0$.
Thus fine-tuning problem accompanied by 
a small scalar mass may be remarkably improved.

On the other hand the anomalous dimension 
makes the quartic coupling 
highly irrelevant, since the scaling dimension
of $\lambda_4$ is given by $4 \gamma_{\phi}^*$
at the IR fixed point.
This does not mean that the quartic coupling
is eliminated.
The RG equation (\ref{quarticRG}) tells us that
$\lambda_4$ approaches the fixed point
valule given by 
\be
\lambda_4^* \simeq
|y^*|^2
= \frac{(4\pi)^2}{2N_c n_f }\gamma_{\phi}^*,
\ee
very strongly. Similarly the couplings of higher
point interactions of the scalar converge to
the fixed point values strongly.
The presence of the quartic coupling in
the RG equation does not alter these
renormalization properties.

Lastly we shall consider renormalization of 
the scalar mass 
in the theories on the renormalized 
trajectory from the BZ fixed point.
In a general scalar theory, hierarchically small mass
cannot be realized, unless
the parameter
$\tilde{m}_{\phi}^2(\Lambda)$ given to the cutoff
theory is finely tuned to a certain non-zero value.
Contrary to this, the CFT of the IR fixed point
of the gauge-Yukawa theory 
may be realized by adding infinitesimally small
Yukawa coupling with a {\it massless free scalar}
to the gauge theory on the BZ fixed point.

In Fig.~9, the RG flows in the
space of $(\alpha_y, \tilde{m}_{\phi}^2)$
obtained by solving Eq.~(\ref{scalarmassRG})
are shown in the case of 
$r=4.1$ and $n_f=1$.
Then the anomalous dimension at the IR fixed point
is found to be $\gamma_{\phi}^* \simeq 0.65$.
The blobs stand for the fixed points
$(\alpha_y^*, \tilde{m}_{\phi}^{2*})$.
At the BZ fixed point, the quartic coupling
$\lambda_4$ is set to zero. 
The figure shows that there is a 
renormalized trajectory connecting these
fixed points. 
The RG flow lines around it show the relatively
slow evolution of the mass parameter,
which is caused by the large anomalous dimension.
Then it is seen that the flows pass vicinity of
the IR fixed point can be obtained without
fine-tuning of $\tilde{m}_{\phi}^2$.
This behavior is very contrasting compared
with the cases with small anomalous dimension.
In Fig.~10, the RG flows near the renormalized 
trajectory towards the IR fixed point are also shown
in the case of $r=5$ and $n_f=1$ for comparison.
In this case the fixed point couplings are rather
small and the anomalous dimension is given by
$ \gamma_{\phi}^* \simeq 0.08$.
It is clearly seen that the RG flows deviate
from the renormalized trajectory very easily
in sharp contrast with the case of $N_f=12$.
Thus a small scalar mass cannot be realized
without tremendous fine-tuning.

\begin{figure}[htb]
\includegraphics[width=0.5\textwidth]{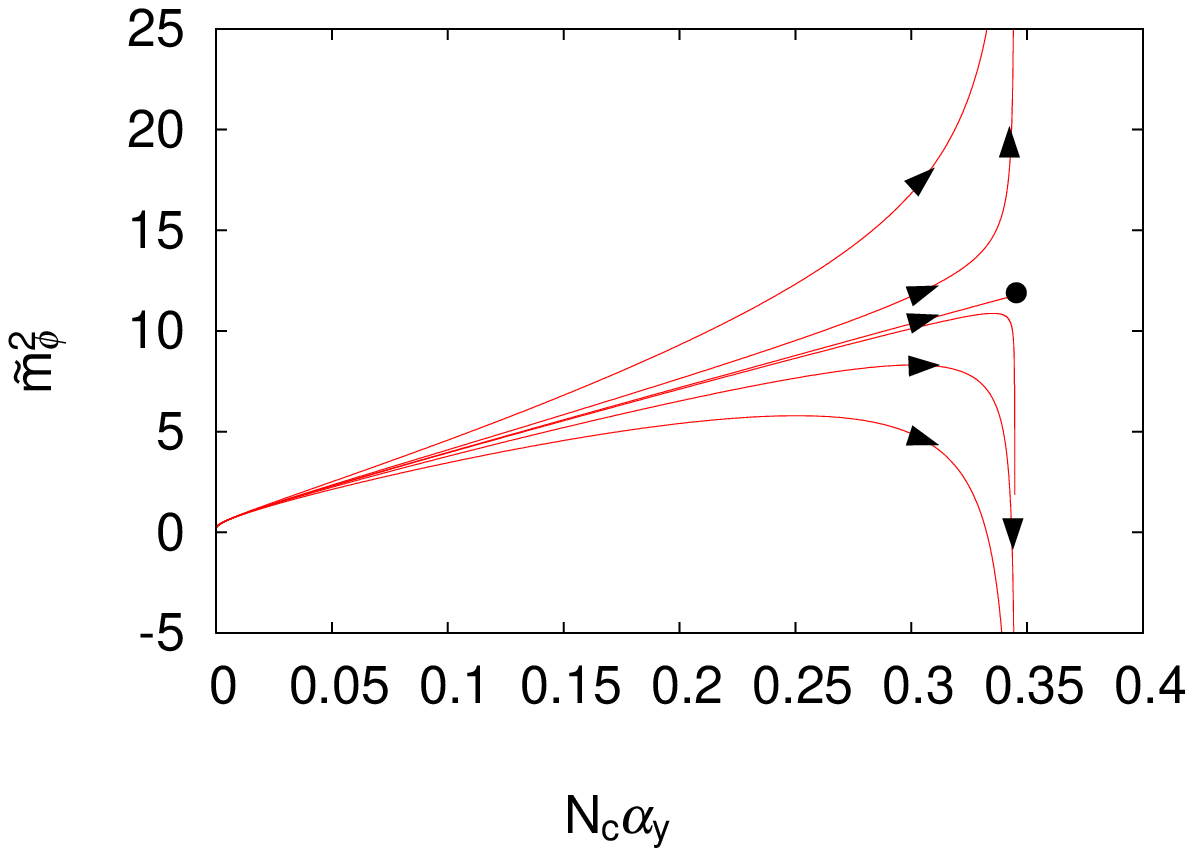}
\caption{\label{fig10}
RG flows of the dimensionless mass parameter
$\tilde{m}_{\phi}^2$ in the case of 
$r=4.1$ and $n_f=1$ are shown in the
space of $(\alpha_y, \tilde{m}_{\phi}^2)$.
The blobs stand for the fixed points
$(\alpha_y^*, \tilde{m}_{\phi}^{2*})$.
}
\includegraphics[width=0.5\textwidth]{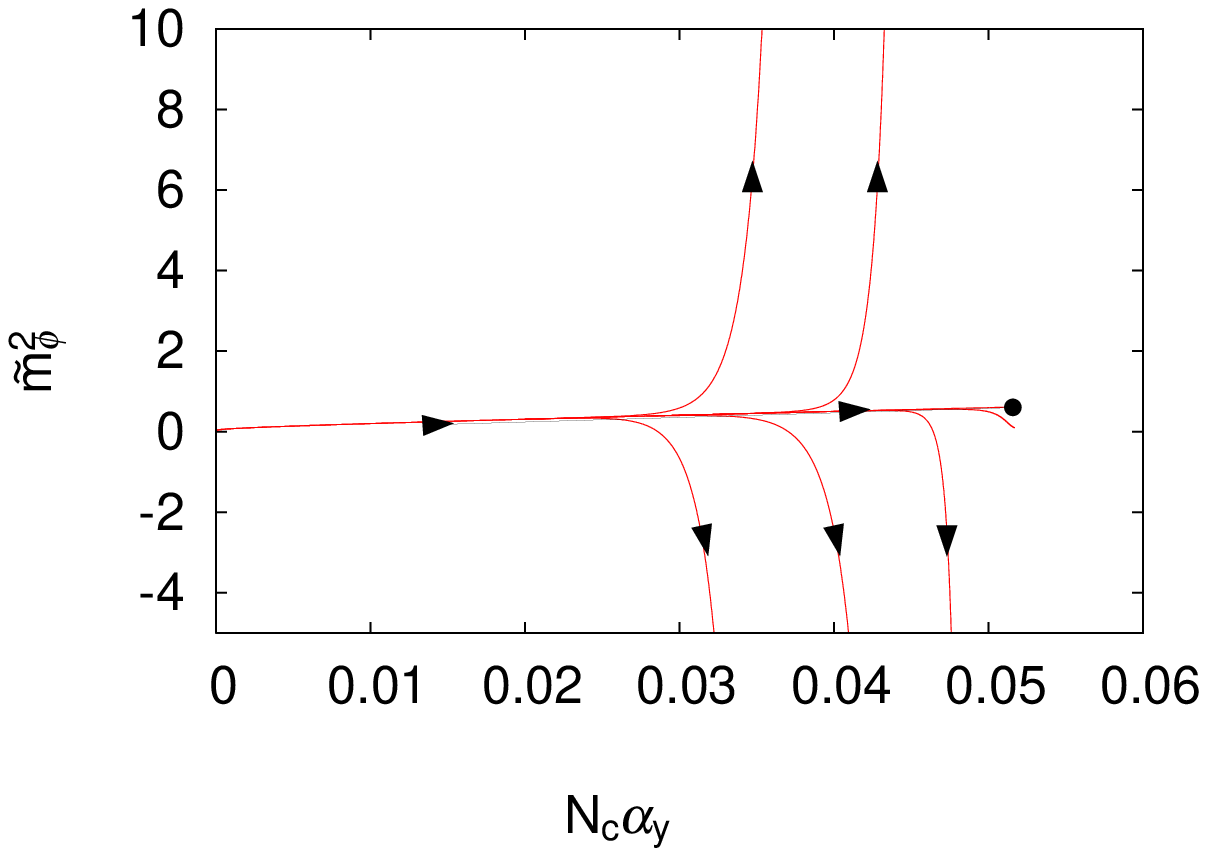}
\caption{\label{fig11}
RG flows of $\tilde{m}_{\phi}^2$ in the case of 
$r=5$ and $n_f=1$ are shown for comparison.
}
\end{figure}

\section{Conclusions and discussions}

We studied the fixed points and the phase
structure of the
gauge-Yukawa theories with massless fermions
by means of the NPRG.
We adopted a somewhat crude approximation
scheme, in which the RG equation for the
gauge coupling is substituted with the two-loop
beta function.
However, it was explicitly shown that the 
approximation beyond the conventional
ladder approximation may be applied
easily owing to the simplicity of the RG framework.
It was also demonstrated through the analysis
that the NPRG approach is superior
to the DS approach in discussing the fixed points
and dynamics around them.

The results of the analysis show that the 
anomalous dimension of the fermion mass operator
can be nearly -1 at the BZ fixed points, 
when the flavor number
is close to the lower edge of the 
conformal window.
Then the Yukawa interaction operator 
with a scalar field is relevant at the fixed point.
We may find that there is the IR fixed point
with non-trivial Yukawa coupling in the gauge-Yukawa
theories with a flavor singlet scalar.
At the IR fixed point, the scalar field acquires
a large positive anomalous dimension.

We also discussed the effect of the large anomalous 
dimension to renormalization of the scalar potential.
Especially, the cutoff dependence of the scalar
mass squared may be drastically reduced.
The other self interactions of the scalar are all 
irrelevant and the couplings converge to their
fixed point values.
The scalar mass vanishes as renormalization scale
goes to zero due to the anomalous dimension
of the fixed point theories.
It is found that the gauge-Yukawa theories on the 
renormalized trajectory give also massless scalars.
Thus the gauge-Yukawa theories controlled by 
the IR fixed point are not annoyed with the 
fine-tuning problem in order
to realize a large mass hierarchy.

In this paper, we have not discussed the dynamics
in the broken phase of gauge theories with many
flavors, although a lot of studies
have been done by the DS approach.
We may evaluate the chiral order parameters 
in the NPRG framework as is shown 
in Ref.~\cite{NPRG}.
Especially it is interesting not only theoretically
but also phenomenologically to evaluate
various order parameters near the critical surface,
or near the UV fixed point.
For example, the S-parameter and the pion decay constant
in the gauge theories with
massless flavors slightly less than the critical
number have been investigated by solving the
Bethe-Salpeter equations as well as the DS equations
recently \cite{brokenphase}.
Such studies seem very interesting in the viewpoint of
the phenomenologically viable model building of 
composite Higgs.
Here we leave these problems to the future works.

\section*{Acknowledgements}
H.T is supported in part by the Grants-in-Aid for 
Scientific Research (No.~14540256) and (No.~16028211).
from the Ministry of Education, Science, Sports and 
Culture, Japan.

\appendix
\section{NPRG equations for the four-fermi
couplings}

We demonstrate the explicit derivations 
of the RG equations for the four-fermi
couplings given by Eq.~(\ref{4fermilagrangian}).
The ERG equation is reduced to a set of infinite
numbers of one-loop RG equations, once we perform
the operator expansion.
Evolution of each coupling may be given by
reducing the scale of momentum cutoff $\Lambda$
infinitesimally.
This is carried out by 
integration of the ``shell'' modes, whose
momentum $p$ belongs to 
$[\Lambda - \delta \Lambda, \Lambda]$,
in evaluating the Wick rotated one-loop diagrams.
Let us start with the one-loop diagrams inducing the
four-fermi couplings, which are 
illustrated in Fig.~12.
We use the Euclidean propagators for the 
fermion $S(p)=i/\psla$ and for the
gauge boson 
\be
D_{\mu \nu}(p) = \frac{1}{p^2}\left(
\delta_{\mu \nu} - (1 - \xi) \frac{p_{\mu} p_{\nu}}{p^2}
\right), 
\ee
where $\xi$ denotes the gauge parameter.

\begin{figure}[htb]
\includegraphics[width=0.5\textwidth]{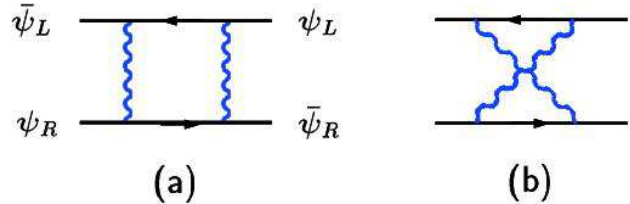}
\caption{\label{fig12}
The four-fermi interaction induced by the gauge
interaction at the one-loop level.}
\end{figure}

The four-fermi operator induced by the
box diagram (a) in Fig.~12, 
which generates the ladder diagrams 
by solving the NPRG equations, 
is evaluated as 
\bea
\delta {\cal L_{\rm eff}} 
&=&
-g^4
\int_{\Lambda-\delta\Lambda < |p|< \Lambda}
\frac{d^4 p}{(2\pi)^4} 
D_{\mu \nu}(p)D_{\rho \sigma}(p)
\nn \\
& & \hspace{10pt}
\sum_{AB}
\bar{\psi}_{Li} T^A \gamma_{\mu}
S(p)T^B \gamma_{\rho}\psi_L^i~
\bar{\psi}_{Rj} T^B \gamma_{\sigma}
S(p) T^A \gamma_{\nu} \psi_R^j
\nn \\
&=& 
\frac{9}{2\Lambda^2} \alpha_g^2 \sum_{AB} 
\bar{\psi}_{Li} T^A T^B \gamma_{\mu} \psi_L^i~
\bar{\psi}_{Rj} T^B T^A \gamma_{\mu} \psi_R^j
~\frac{\delta \Lambda}{\Lambda}.
\eea
This operator may be transformed to the operator
${\cal O}_S$ given by (\ref{OS})
by using  the Fierz identity, 
\be
\sum_{AB}
\bar{\psi}_{Li} T^A T^B \gamma_{\mu} \psi_L^i~
\bar{\psi}_{Rj} T^B T^A \gamma_{\mu} \psi_R^j
= - \frac{2}{N_c} C_2(N_c)^2 
\bar{\psi}_{Li}\psi_R^j~
\bar{\psi}_{Rj}\psi_L^i.
\ee
Eventually we may obtain a part of the RG equation for the
four-fermi coupling $g_S$ as
\be
\Lambda\frac{dg_S}{d\Lambda} = 18 \frac{C_2(N_c)^2}{N_c}
\alpha_g^2.
\ee

In the same way the crossed box diagram (b)
in Fig.~12 generates 
\be
\Lambda\frac{dg_S}{d\Lambda} =
6 \frac{C_2(N_c)}{N_c}
\left(
C_2(N_c) - \frac{N_c}{2}
\right) \alpha_g^2.
\ee
This contribution is sub-leading in the large
$N_c$ expansion.

It has been found that the summation
of these diagrams give a gauge independent 
corrections for abelian gauge theories \cite{NPRG}.
In the non-abelian gauge theories, however,
sum of these corrections is gauge depend.
In order to  make it gauge independent, it would be
necessary to incorporate higher dimensional
operators such as
$\bar{\psi}\sigma_{\mu \nu}\psi F^{\mu \nu}$.
We leave this extension for the future study.
Rather, in this paper we adopt the Landau gauge.

The other diagrams can be evaluated in a similar
manner.
Thus the RG equations for the four-fermi couplings
given by (\ref{4fermilagrangian})
are found to be
\bea
\mu \frac{d g_S}{d \mu}
&=& 2 g_S
- 2 N_c g_S^2 + 2 N_f g_S g_V + 6 g_S g_{V1}
+ 2 g_S g_{V2} \nn \\
& &
-12 C_2(N_c)\alpha_g g_S 
- 18 \frac{C_2(N_c)^2}{N_c} \alpha_g^2
- 6 \frac{C_2(N_c)}{N_c}
\left(
C_2(N_c) - \frac{N_c}{2}
\right) \alpha_g^2, 
\label{gSRG} \\
\mu \frac{d g_V}{d \mu}
&=& 2 g_V + \frac{1}{4} N_f g_S^2 + (N_c + N_f)g_V^2
- 8 g_V g_{V2} \nn \\
& &
-3 \frac{C_2(N_c)^2}{N_c} \alpha_g^2
- 9  \frac{C_2(N_c)}{N_c}
\left(
C_2(N_c) - \frac{N_c}{2}
\right) \alpha_g^2, 
\label{gVRG} \\
\mu \frac{d g_{V1}}{d \mu}
&=& 2 g_{V1} -\frac{1}{4}g_S^2 - g_S g_V
-N_f g_S g_{V2} + 2(N_c-N_f)g_V g_{V1}
\nn \\
& &
+ 3 g_{V1}^2 + 2(N_c N_f +1 )g_{V1}g_{V2}
-\frac{1}{2}  \frac{C_2(N_c)}{N_c} \alpha_g g_S,
\label{gV1RG} \\
\mu \frac{d g_{V2}}{d \mu}
&=& 2 g_{V2} - 3g_V^2 -(N_f + 2) g_S g_{V1}
+ (2N_c + 4 N_f - 8)g_V g_{V2} 
\nn \\
& &
+ N_f g_{V1}^2 + (N_c N_f -2 )g_{V2}^2,
\label{gV2RG}
\eea
where $g_A = G_A/(4\pi^2)$ $(A = S, V, V1, V2)$. 
The Eqs.~(\ref{largeNGSRG}) and (\ref{largeNGVRG})
are extracted from the above equations in the
large $N_c$ and $N_f$ leading.

\end{document}